\documentclass{elsart}

\usepackage{graphicx}

\usepackage{amssymb}

\begin{document}

\begin{frontmatter}



\title{Design of beam splitters and microlasers using chaotic
waveguides}

\author[MPI]{O. Bendix},
\author[MPI]{J. A. M\'endez-Berm\'udez\corauthref{cor}\thanksref{GIF}},
\author[IFUAP]{G. A. Luna-Acosta\thanksref{Merc}},
\author[MAR]{U. Kuhl},
\author[MAR]{H. -J. St\"{o}ckmann}
\corauth[cor]{Corresponding author: antonio@chaos.gwdg.de}
\address[MPI]{Max-Planck-Institut f\"ur Dynamik und Selbstorganisation, Bunsenstra\ss e 10,
D-37073 G\"ottingen, Germany}
\address[IFUAP]{Instituto de
F\'{\i}sica, Universidad Aut\'onoma de Puebla, Apartado Postal
J-48, Puebla 72570, M\'exico}
\address[MAR]{Fachbereich Physik der Philipps Universit\"at Marburg, Renthof 5, D-35032
Marburg, Germany}
\thanks[GIF]{Supported by the GIF, the German-Israeli Foundation
for Scientific Research and Development.}
\thanks[Merc]{Partially Supported by Mercator Professorship, Germany.}

\begin{abstract}
We consider waveguides formed by single or multiple
two-dimensional chaotic cavities connected to leads. The cavities
are chaotic in the sense that the ray (or equivalently, classical
particle) dynamics within them is chaotic. Geometrical parameters
are chosen to produce a mixed phase space (chaotic regions
surrounding islands of stability where motion is regular).
Incoming rays (or particles) cannot penetrate into these islands
but incoming plane waves dynamically tunnel into them at a certain
discrete set of frequencies (energies). The support of the
corresponding quasi-bound states is along the trajectories of
periodic orbits trapped within the cavity. We take advantage of
this difference in the ray/wave behavior to demonstrate how
chaotic waveguides can be used to design beam splitters and
microlasers. We also present some preliminary experimental results
in a microwave realization of such chaotic waveguide.
\end{abstract}

\end{frontmatter}

\section{Introduction}

During the last years there has been a growing interest in the use
of chaotic cavities as resonators for microlasers, mainly due to
the possibility of quality factor tuning and its highly
directional emission, see for example \cite{GCNNSFSC}. The key
feature behind these characteristics is the cavity geometry, which
is designed in most cases to produce mixed chaotic dynamics.

Here we propose a two-dimensional (2D) locally deformed waveguide
that can be used as beam splitter or as (single or multi-cavity)
resonator for microlasers. The novel and important feature of our
model is that the cavities are open, unlike the well known model
of Ref. 1. This difference allows for a wider versatility and
range of applications. The splitting and lasing mechanism requires
that the deformation yields particle (ray) motion in a mixed phase
space. We remark that  due to the equivalence between the problem
of a TM wave inside a 2D waveguide with Dirichlet boundary
conditions (Helmholtz equation) to that of a quantum wave in a 2D
billiard (Schr\"{o}dinger equation) \cite{stock}, our results are
applicable to electromagnetic as well as electronic setups.

\section{The Model}

\begin{figure}
\begin{center}
\includegraphics*[width=10cm]{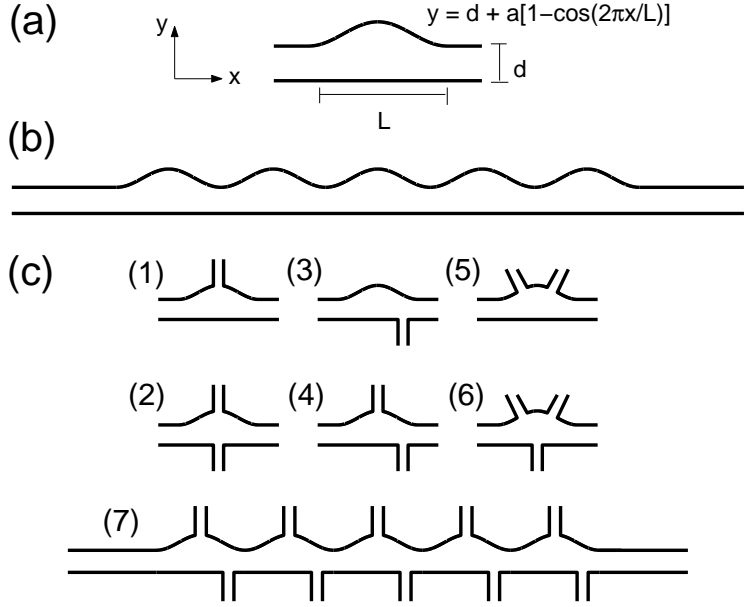}
\end{center}
\caption{(a) Geometry of the cavity, (b) a multi-cavity waveguide,
and (c) examples of beam splitter setups.}
\label{fig:setups}
\end{figure}

The 2D waveguide we shall use consists of a cavity connected to
two collinear semi-infinite leads of width $d$ extended along the
$x$-axis. The prototype cavity has the geometry of the so-called
{\it cosine billiard} extensively studied in
\cite{infinite:s,finite2}: it has a flat wall at $y=0$ and a
deformed wall given by $y(x)=d+a [1-\cos(2\pi x/L)]$, where $a$ is
the amplitude of the deformation and $L$ is the length of the
cavity. In Fig. \ref{fig:setups}(a) we show the geometry of the
cavity.

\begin{figure}
\begin{center}
\includegraphics*[width=8cm]{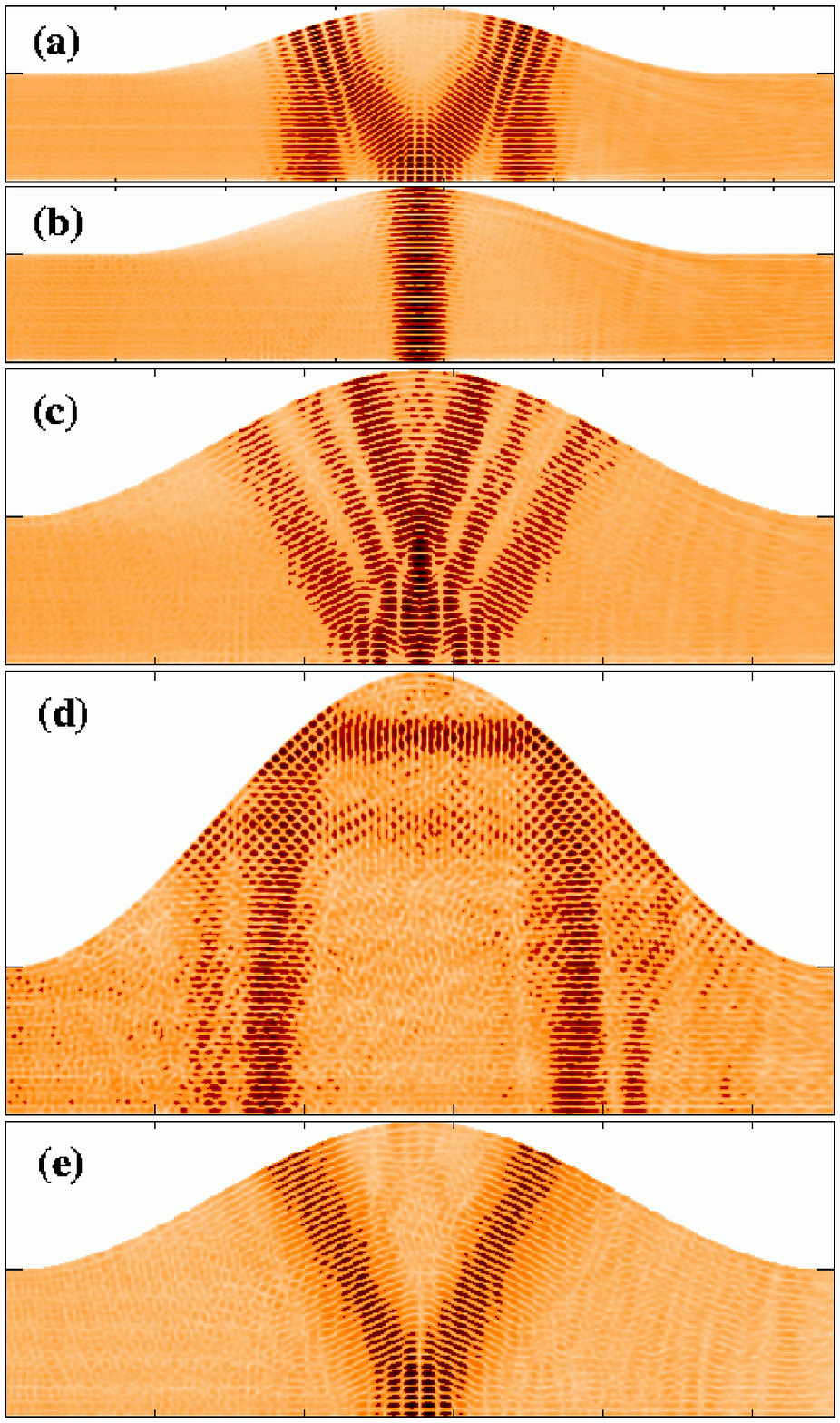}
\end{center}
\caption{Wave function density plots for the waveguide of Fig.
\ref{fig:setups}(a), showing (a) M-, (b) I-, (c) W-, (d) $\Pi$-,
and (e) V-type quasi-bound states.
See \cite{finite2,NOVA,MLSP03} for the calculation details.}
\label{fig:QBS}
\end{figure}

For any $a>0$ the cavity develops a ternary horseshoe
\cite{Wiggins} that can be either incomplete (proper of mixed
chaotic dynamics) or complete (prescribing global chaos),
depending on the particular values of $(d,a,L)$.\cite{NOVA} When a
waveguide is constructed with a cavity characterized by an
incomplete horseshoe, its conductance fluctuates strongly with
sharp resonances \cite{BHK02,finite2}.The wave functions
corresponding to the {\it sharpest} conductance resonances can be
identified with energy eigenstates living in phase space stability
islands, whence  they are  {\it quasi bound states} (QBS)
\cite{finite2}. Note that incoming trajectories cannot penetrate
into the resonance (stability) island as motion in the islands
correspond to trajectories trapped within the cavity. However, the
wave function dynamically tunnels into the islands at the
resonance values, allowed by Heisenberg's uncertainty principle.
All QBS have support on stability islands surrounding low period
periodic orbits. In particular, for $(d,a,L)=(1.0,0.305,5.55)$ the
QBS reveal I- and M-shaped patterns while for
$(d,a,L)=(1.0,0.5,5.55)$ V- and W-shaped patterns are observed. In
Fig. \ref{fig:QBS} examples of I-, M-, V-, and W-type QBS are
shown.

\begin{figure}
\begin{center}
\includegraphics*[width=10cm]{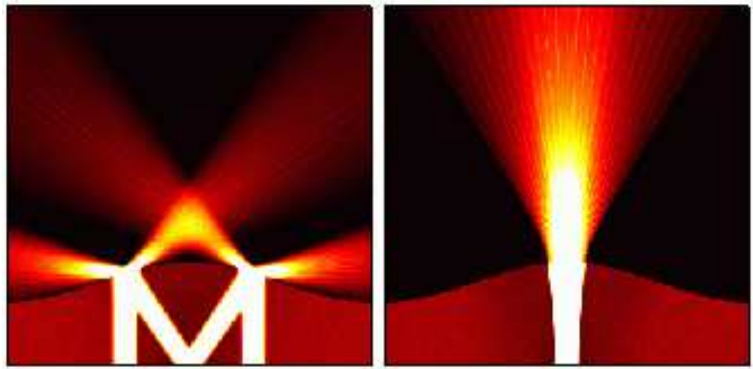}
\end{center}
\caption{Ray prediction of lasing emission for M- and I-type quasi-bound states.}
\label{fig:laser}
\end{figure}

\begin{figure}
\begin{center}
\includegraphics*[width=12cm]{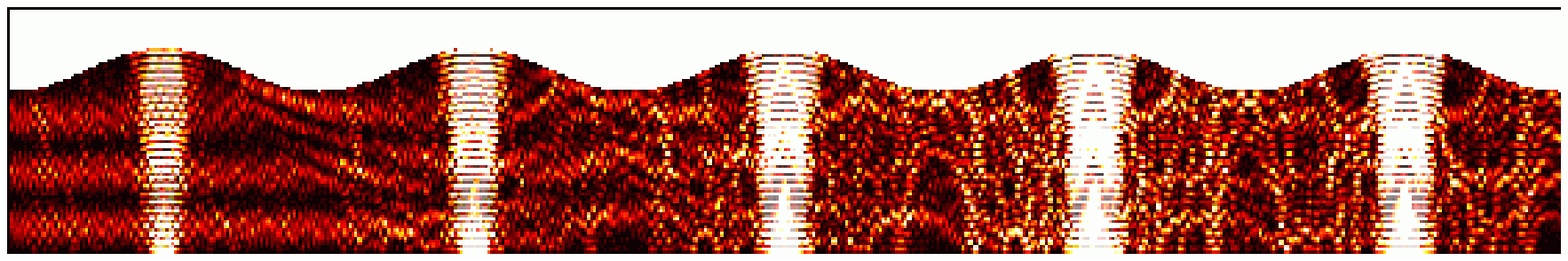}
\end{center}
\caption{Wave function density plots for the multi-cavity waveguide of Fig.
\ref{fig:setups}(b) showing I-type quasi-bound states.}
\label{fig:multic}
\end{figure}

\section{Discussion}

\subsection{Microlasers}

If the waveguide described above is constructed with a
semiconductor material having a refraction index $n$ and a wave is
introduced into one of the leads with an energy corresponding to a
sharp conductance resonance, the cavity acts as a laser resonator
since light will bounce in the cavity region having the chance to
escape only for certain incidence angles, which depend on the type
of the QBS excited and the index of refraction of the waveguide.
In Fig. \ref{fig:laser} ray dynamics was used to predict the
direction and intensity of the lasing produced by M- and I-type
QBS for typical values of refraction index, showing both high
quality and high directionality \cite{MLSP03}. The great advantage
of a waveguide-based resonator may be in that no pedestals or
couplers (pumpers) close to the cavity are needed, as is the case
with the currently investigated 3D and 2D micro lasers.

\subsection{Multi-cavity resonators}

Our waveguide model can also be used to construct multi-cavity
resonators formed, for example, by a co-linear array of coupled 2D
chaotic cavities, c.f. Fig. \ref{fig:setups}(b). Here, as the
number of cavities is increased most of the ray trajectories take
longer to be transmitted or reflected; they oscillate irregularly
around the widest parts of the coupled cavities. The dwelling of
the rays in the outskirts of the resonance islands induces a
higher probability of dynamical tunneling into the classical
inaccessible regions. Thus, the trapping of scattering wave
functions along bounded ray trajectories is enhanced. See in Fig.
\ref{fig:multic} an I-type QBS excited in the milti-cavity
resonator of Fig. \ref{fig:setups}(b). Even though a multi-cavity
resonators is of general interest in the field of optics, one may
also consider the construction of a micro laser using a resonant
multi-cavity. A multi-cavity micro laser would have two main
advantages: (i) the multi-cavity array would enhance considerably
the emission intensity compared to a micro laser constructed with
one cavity only, and (ii) the micro laser would emit several
parallel beams in one or more directions since I-, M-, V-, W-, or
$\Pi$-type QBS can be excited by choosing the appropriate cavity
geometry \cite{OML}.

\subsection{Beam splitters}

\begin{figure}
\begin{center}
\includegraphics*[width=8cm]{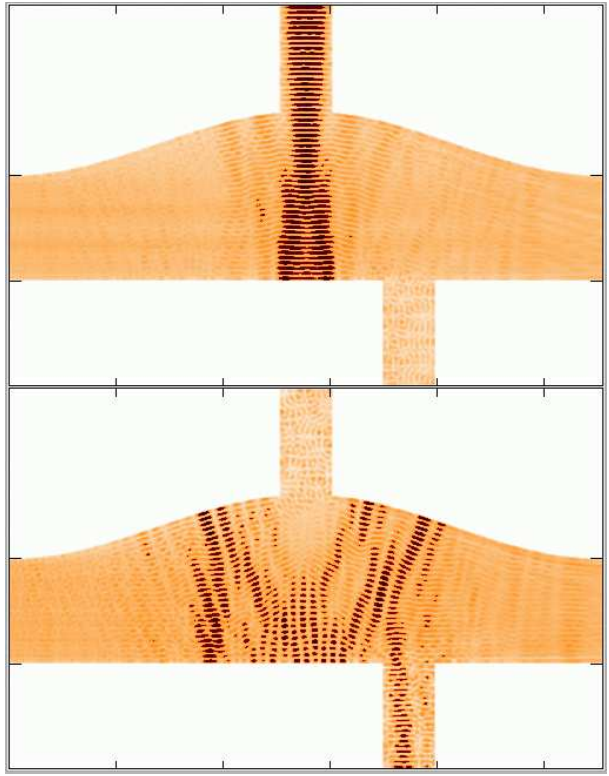}
\end{center}
\caption{Wave function density plots for two examples of beam
splitters.} \label{fig:splitter}
\end{figure}

Once one knows the type of QBS that can be excited for a given set
of cavity parameters $(d,a,L)$, it is possible to construct
electromagnetic or electronic beam splitters by attaching
transversal leads to the waveguide. The transversal leads are
prescribed as follows \cite{OML}: (i) they have to be located {\it
on} the stability islands supporting the QBS, and (ii) their width
must be small enough to preserve the global phase space structure.
For example, Figs. \ref{fig:setups}(c1-c2), (c3), and (c5-c6) show
beam splitter setups suitable for I-, M-, and V-type QBS,
respectively. We still consider plane waves coming from the
horizontal leads. Then, the scattering wave functions which tunnel
into stability islands will be guided out of the cavity through
the transversal leads. Fig. \ref{fig:setups}(c4) is an example of
a beam splitter using both I- and M-type QBS, where the beam will
be guided up or down depending on the excited QBS, as shown in
Fig. \ref{fig:splitter}. Obviously, one can also anticipate the
construction of multi beam splitters, as suggested in Fig.
\ref{fig:setups}(c7).

\subsection{Experimental}

\begin{figure}
\begin{center}
\includegraphics*[width=14cm]{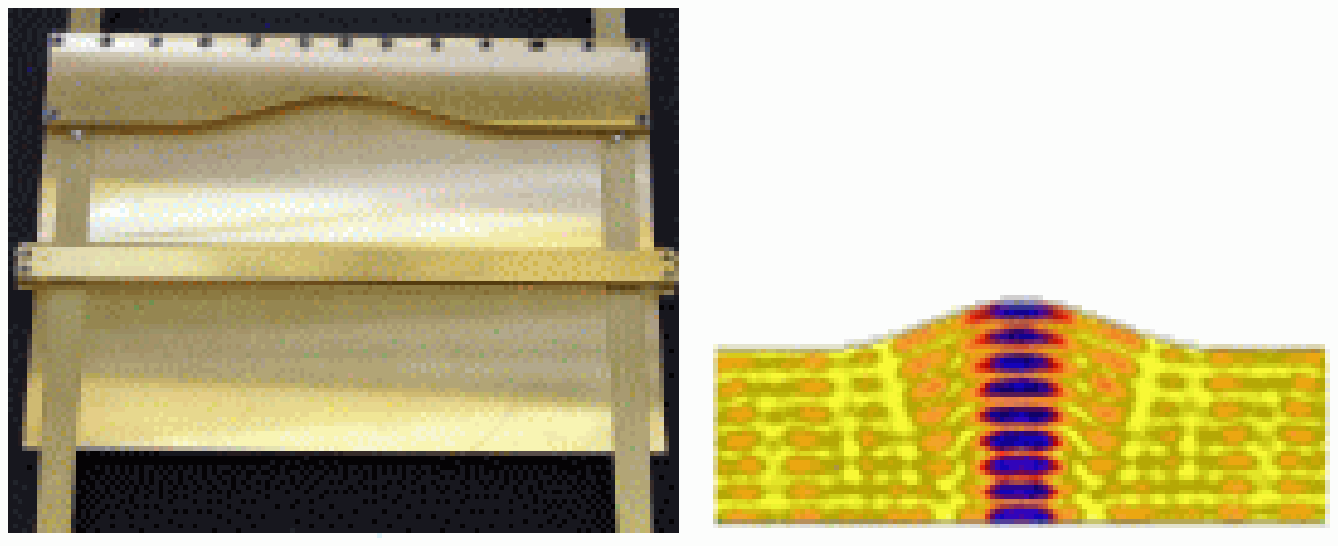}
\end{center}
\caption{Left: Microwave Cosine billiard. Right: I-type QBS,
experimentally obtained.}
\label{fig:expwf}
\end{figure}

We have recently performed measurements on the experimental
realization of the cosine billiard in the microwave regime. Figure
\ref{fig:expwf} shows the experimental setup. Not shown in the
figure is a top metallic plate. The distance between the top and
bottom plates, the resonator height $h$, is 0.8 cm. As long as the
frequency $\nu$ is less than $c/2h=18.75$ GHz ($c$ is the speed of
light) the cavity is a 2D system and there is a one to one
correspondence between the electric field strength and the quantum
mechanical wave function (see \cite{stock}). The present set up
includes three electric dipole antennas. The fist one is fixed on
the left lead, near the top left corner, at $(x,y)=(-18.0,7.725)$
cm, the second one is fixed on the opposite side, at $(x,y)=(18.0,
7.725)$ cm. The third antenna is inside the cavity, scanning in
small steps the whole cavity region (normally in steps of 2.5 mm)
for each value of frequency in the range 1-18.75 GHz. A vector
network analyzer is used to measure reflection in each of the
antennas as well as transmission between them. Using the methods
developed in \cite{stock2:s} the modulus and sign of the wave
function can be obtained from the transmission measurements, while
the reflection measurement provides only the modulus of the wave
function. In Fig. \ref{fig:expwf} we present a wave function
obtained from the reflection measurements from the scanning
antenna at $\nu=13.7708$ GHz, a resonance frequency. Experimental
results like this confirm the existence of I-type QBS,
corresponding to the theoretical predictions of Fig.
\ref{fig:QBS}(b), and forecast the realization of the proposed
waveguide-based beam splitters and (multi-cavity) resonators.

Finally, as an additional note we want to mention that in the
case of a waveguide with a cavity characterized by a complete
horseshoe the conductance is in general a smooth function of the
energy (or frequency). However, some non-generic complete
horseshoes anticipate the appearance of {\it scars} \cite{Heller},
which in turn produce conductance resonances and $\Pi$-type QBS \cite{OML}.
See an example of a $\Pi$-type QBS in Fig. \ref{fig:QBS}(d) where
$(d,a,L)=(1.0,1.0,5.55)$ was used. Then the construction of
microlasers and beam splitters based on scared QBS could be also
accomplished.

\end{document}